\documentclass[11pt]{article}
\usepackage{amsmath,amssymb,epsf}

\let\X=\Xi

\def\nn{\nonumber}

\let\bm=\bibitem

\newcommand{\be}{\begin{equation}}
\newcommand{\ee}{\end{equation}}
\def\ba{\begin{array}}
\def\ea{\end{array}}
\def\ft#1#2{{\textstyle{\frac{\scriptstyle #1}{\scriptstyle #2}}}}
\def\fft#1#2{\frac{#1}{#2}}
\def\del{\partial}

\def\sst#1{{\scriptscriptstyle #1}}

\def\td{\tilde}
\def\wtd{\widetilde}
\def\ie{\rm i.e.\ }
\def\dalemb#1#2{{\vbox{\hrule height .#2pt
        \hbox{\vrule width.#2pt height#1pt \kern#1pt
                \vrule width.#2pt}
        \hrule height.#2pt}}}

\newcommand{\hoch}[1]{$\, ^{#1}$}
\newcommand{\bea}{\begin{eqnarray}}
\newcommand{\eea}{\end{eqnarray}}

\def\0{{\sst{(0)}}}
\def\1{{\sst{(1)}}}
\def\2{{\sst{(2)}}}
\def\3{{\sst{(3)}}}
\def\4{{\sst{(4)}}}
\def\5{{\sst{(5)}}}
\def\6{{\sst{(6)}}}
\def\7{{\sst{(7)}}}
\def\8{{\sst{(8)}}}

\def\cA{{{\cal A}}}

\def\im{{{\rm i}}}

\def\ep{{\epsilon}}

\def\R{\rlap{\rm I}\mkern3mu{\rm R}}

\def\R{\rlap{\rm I}\mkern3mu{\rm R}}

\def\R{{{\mathbb R}}}

\def\CP{{{\mathbb C}{\mathbb P}}}

\def\bog{{Bogomol'nyi\ }}

\begin{document}
\begin{flushright}
MIFP-06-01 \\
{\bf hep-th/0601002}\\
January\  2006
\end{flushright}

\begin{center}

{\Large {\bf Kerr-de Sitter Black Holes with NUT Charges}}

\vspace{20pt}

W. Chen, H. L\"u and C.N. Pope

\vspace{20pt}

{\hoch{\dagger}\it George P. \&  Cynthia W. Mitchell Institute
for Fundamental Physics,\\
Texas A\&M University, College Station, TX 77843-4242, USA}

\vspace{40pt}

\underline{ABSTRACT}
\end{center}

   The four-dimensional Kerr-de Sitter and Kerr-AdS black hole metrics
have cohomogeneity 2, and they admit a generalisation in which an
additional parameter characterising a NUT charge is included.  In this
paper, we study the higher-dimensional Kerr-AdS metrics, specialised
to cohomogeneity 2 by appropriate restrictions on their rotation
parameters, and we show how they too admit a generalisation in which
an additional NUT-type parameter is introduced.  We discuss also the
supersymmetric limits of the new metrics.  If one performs a Wick
rotation to Euclidean spacetime signature, these yield new
Einstein-Sasaki metrics in odd dimensions, and Ricci-flat metrics in
even dimensions.  We also study the five-dimensional Kerr-AdS black
holes in detail.  Although in this particular case the NUT parameter
is trivial, our investigation reveals the remarkable feature that a
five-dimensional Kerr-AdS ``over-rotating'' metric is equivalent,
after performing a coordinate transformation, to an under-rotating
Kerr-AdS metric.

{\vfill\leftline{}\vfill \vskip 10pt \footnoterule {\footnotesize
Research supported in part by DOE grant
DE-FG03-95ER40917.
}


\newpage
\tableofcontents
\addtocontents{toc}{\protect\setcounter{tocdepth}{2}}

\section{Introduction}

    The discovery by Kerr in 1963 \cite{kerr} of the exact metric
describing a rotating black hole was arguably the most important
advance in the study of exact solutions in general relativity since
Schwarzschild's discovery in 1916 \cite{schw} of the metric describing
a static black hole.  It was followed within a few years by the
finding of various generalisations, including charged rotating black
holes \cite{newman}, and then the further inclusion of a cosmological
constant \cite{carter}, NUT parameter \cite{pleb} and acceleration
parameter \cite{plebdem}.

    With the advent of supergravity and superstring theory, interest
also developed in studying higher-dimensional solutions of the
Einstein equations.  In 1986, the general solution describing an
asymptotically flat rotating black hole in arbitrary dimension was
discovered by Myers and Perry \cite{myeper}.  In dimension $D$, this
has a mass parameter and $[(D-1)/2]$ independent rotation parameters,
one for each of the orthogonal spatial 2-planes.  Motivated by the
study of the AdS/CFT correspondence in string theory, Hawking, Hunter
and Taylor-Robinson constructed the solution for the five-dimensional
rotating black hole with a cosmological constant in 1998
\cite{hawhuntay}.  They also obtained a special case of the metrics in
all higher dimensions, in which there is a rotation in only one of the
$[(D-1)/2]$ orthogonal spatial 2-planes.  The general rotating Kerr-de
Sitter black hole solution with a cosmological constant in arbitrary
dimension, with all $[(D-1)/2]$ independent rotation parameters $a_i$,
was constructed by Gibbons, L\"u, Page and Pope in 2004
\cite{gilupapo1,gilupapo2}.

    In view of the fact that the four-dimensional rotating black hole
metrics admit further generalisations where additional non-trivial
parameters are present, one might wonder whether such additional
parameters could also be introduced in higher dimensions too.  In
fact, in a certain special class of higher-dimensional Kerr-de Sitter
black holes, namely those in which there is just a rotation in a
single 2-plane, a generalisation which includes a NUT parameter as
well as the mass and the (single) rotation parameter has been obtained
\cite{klemm}.  It was shown in \cite{chgilupo} that this
generalisation, which is trivial in five dimensions but non-trivial in
dimensions $D\ge 6$, still exhibits certain remarkable separability
properties for the Hamilton-Jacobi and wave equations, which in fact
played an important r\^ole in the original discovery of the
generalised four-dimensional solutions.

   The purpose of this paper is to present new results we have
obtained for much wider classes of generalisations of the Kerr-de
Sitter metrics, in which there is a NUT-type parameter as well as the
mass parameter.  The cases covered by our new solutions are when the
rotation parameters $a_i$ are divided into two sets, in which all
parameters within a set are equal.  In odd dimensions, which we
discuss in section 2, we obtain generalised solutions for an arbitrary
partition of the parameters into two such sets.  In even dimensions,
which we discuss in section 3, the parameters are partitioned into one
set with a non-vanishing value for the rotation, and the other set
with vanishing rotation.  In each of the odd and even dimensional
cases, the net effect is to give a metric of cohomogeneity 2.  In a
manner that parallels rather closely the generalisations in $D=4$, the
two associated coordinates, on which the metric functions are
intrinsically dependent, enter in a rather symmetrical way. The
metrics that we obtain are equivalent to the previously-known Kerr-de
Sitter-Taub-NUT metrics in $D=4$.  In $D\ge6$ the extra parameter that
we introduce gives rise to non-trivial generalisations of the Kerr-de
Sitter metrics.  The new parameter is associated with characteristics
that generalise those of Taub-NUT like metrics in four dimensions, and
so we may think of it as being a higher-dimensional generalisation of
the NUT parameter.  In each of the odd and even-dimensional cases, we
discuss also their supersymmetric limits.  In odd dimensions, these
yield, after Euclideanisation, new Einstein-Sasaki metrics.  In even
dimensions, the supersymmetric limit leads to new Rici-flat K\"ahler
metrics.

   In section 4, we discuss some global aspects of the new
Kerr-AdS-Taub-NUT metrics.  In particular, in the case of even
dimensions, the introduction of the NUT-type parameter implies that
the time coordinate must be identified periodically, in the same way
as happens in the previously-known four-dimensional solutions.  By
contrast, we find that in odd dimensions one can define a time
coordinate that is not periodic.

   In section 5, we discuss the case of five dimensions in detail.  We
find that in this case, the new NUT-type parameter is actually bogus,
in the sense that it can be removed by using a scaling symmetry that
is specific to the five-dimensional metric.  In the process of showing
this, however, we uncover an intriguing and previously unnoticed
property of the five-dimensional Kerr-AdS metric.  We find that it has
an ``inversion symmetry,'' which implies that the metric with large
values of its rotation parameters is equivalent, after a general
coordinate transformation, to the metric with small values for the
rotations.  The fixed point of this symmetry occurs at the critical
value of rotation that arises in the supersymmetric limit.  This
corresponds to the case where the rotation parameter is equal to the
radius of the asymptotically AdS spacetime. The inversion symmetry is
therefore a feature specifically of the five-dimensional Kerr black
holes with a cosmological constant, and does not arise in the case of
asymptotically flat black holes.

   The paper ends with conclusions in section 6.

\section{Kerr-de Sitter with NUT Parameter in $D = 2n + 1 $}
\label{oddsec}
\subsection{The Metric}

   We take as our starting point the general Kerr-de Sitter metric in
$D=2n +1$ dimensions, which was constructed in
\cite{gilupapo1,gilupapo2}.  Specifically, we begin with the metrics
written in an asymptotically non-rotating frame, as given in equation
(E.3) of \cite{gilupapo1}, specialised to the case of odd dimensions
$D=2n+1$.  We choose the cosmological constant to be negative, with
the Ricci tensor given by $R_{\mu\nu}= - (D-1) g^2\, g_{\mu\nu}$.  The
constant $g$ is the inverse of the AdS radius.  The metric is
described in terms of $n$ ``latitude'' or direction cosine coordinates
$\mu_i$, subject to the constraint $\sum_{i=1}^n \mu_i^2=1$, $n$
azimuthal coordinates $\phi_i$, the radial coordinate $r$ and time
coordinate $t$.  It has $(n+1)$ arbitrary parameters $M$ and $a_i$,
which can be thought of as characterising the mass and the $n$ angular
momenta in the $n$ orthogonal spatial 2-planes.

   In order to find a generalisation that includes a NUT-type
parameter, we first specialise the Kerr-AdS metrics by setting
\be
a_1 = a_2= \cdots = a_p = a\,,\qquad
a_{p+1} = a_{p+2}= \cdots = a_{n} = b\,.\label{oddab}
\ee
  We then reparameterise the latitude
coordinates coordinates as
\bea 
\mu_i &=& \nu_i\, \sin\theta\,, \qquad 1\le i\le p\,,\qquad
   \sum_{i=1}^p \nu_i^2=1\,,\nn\\
\mu_{j+p} &=& \td\nu_j\, \cos\theta\,,\qquad 1\le j \le q\,,\qquad
  \sum_{j=1}^q \td\nu_j^2=1\,,
\eea
where we have defined 
\be
n=p+q\,,
\ee
and we also then introduce a coordinate $v$ in place of $\theta$,
defined by
\be
a^2 \cos^2\theta + b^2 \sin^2\theta = v^2\,.\label{thetav}
\ee
It is convenient to divide the original $n$ azimuthal coordinates
$\phi_i$ into two sets, with $p$ of them denoted by $\phi_i$ and the
remaining $q$ denoted by $\td\phi_j$.

   Because of the specialisation of the rotation parameters in
(\ref{oddab}), the Kerr-AdS metric will now have cohomogeneity 2,
rather than the cohomogeneity $n$ of the general $(2n+1)$-dimensional
Kerr-AdS metrics.  In fact, as we shall see explicitly below, the
metric has homogeneous level sets $\R\times S^{2p-1}\times S^{2q-1}$,
with the metric functions depending inhomogeneously on the coordinates
$r$ and $v$.  Remarkably, the form in which the metric can now be
written puts the radial coordinate $r$ and the coordinate $v$ on a
parallel footing, and suggests a rather natural generalisation in
which a NUT-type parameter $L$ can be introduced.  Rather than writing
the metric first without the NUT contribution and then again with it
added, we shall just directly present our final result with the NUT
parameter included.  The original Kerr-AdS, subject to the constraints
on the rotation parameters specified in (\ref{oddab}), corresponds to
setting $L=0$.  Our generalised metric including $L$ is
\bea
ds^2 &=& - \fft{(1+g^2r^2)(1-g^2v^2)}{\Xi_a\Xi_b} dt^2
+\fft{\rho^{2n-2} dr^2}{U} + \fft{\omega^{2n-2} dv^2}{V}\nn\\
&&
+\fft{2M}{\rho^{2n-2}} 
  \Big(\fft{(1-g^2 v^2)}{\Xi_a \Xi_b}\, dt - \cA\Big)^2 +
\fft{2L}{\omega^{2n-2}} 
   \Big(\fft{(1+g^2 r^2)}{\Xi_a\Xi_b}\, dt - \wtd\cA\Big)^2
\label{oddmet}\\
&&
+
\fft{(r^2 + a^2)(a^2-v^2)}{\Xi_a (a^2-b^2)}
\sum_{i=1}^p \Big(d\nu_i^2 + \nu_i^2 d\phi_i^2\Big)+ 
\fft{(r^2 + b^2)(b^2-v^2)}{\Xi_b (b^2-a^2)}
\sum_{j=1}^q \Big(d\td\nu_j^2 + \td\nu_j^2 d\td\phi_j^2\Big)\,,\nn
\eea
where
\bea
\cA&=&\fft{a(a^2-v^2)}{\Xi_a(a^2-b^2)} \sum_{i=1}^p
\nu_i^2 d\phi_i +\fft{b(b^2-v^2)}{\Xi_b(b^2-a^2)} \sum_{j=1}^q
\td\nu_j^2 d\td\phi_j\,,
\nn\\
\wtd\cA&=&\fft{a(r^2+a^2)}{\Xi_a(a^2-b^2)} \sum_{i=1}^p
\nu_i^2 d\phi_i +\fft{b(r^2+b^2)}{\Xi_b(b^2-a^2)} \sum_{j=1}^q
\td\nu_j^2 d\td\phi_j\,,
\nn\\
U&=&\fft{(1 + g^2 r^2)(r^2 + a^2)^p\, (r^2 + b^2)^{q}}{r^2} - 2M\,,
\nn\\
V&=&-\fft{(1-g^2v^2)(a^2-v^2)^p\, (b^2-v^2)^{q}}{v^2} + 2L\,.
\nn\\
\rho^{2n-2}&=& (r^2 + v^2) (r^2 + a^2)^{p-1} \, (r^2 + b^2)^{q-1}
\,,\qquad \Xi_a=1-a^2g^2\,,\nn\\
\omega^{2n-2}&=& (r^2 + v^2) (a^2-v^2)^{p-1}\, 
(b^2-v^2)^{q-1}\,,\qquad\Xi_b=1-b^2 g^2\,.
\eea
It is straightforward (with the aid of a computer) to verify in a
variety of low odd dimensions that the metric (\ref{oddmet}) does
indeed solve the Einstein equations $R_{\mu\nu}= -(D-1) g^2
g_{\mu\nu}$, and since the construction does not exploit any special
features of the low dimensions, one can be confident that the solution
is valid in all odd dimensions.  We have explicitly verified the
solutions in $D\le 9$.

   As we indicated above, the metric (\ref{oddmet}) can be
re-expressed more elegantly in terms of two complex projective spaces
$\CP^{p-1}$ and $\CP^{q-1}$.  The proof is straightforward, following
the same steps as were used in \cite{gilupapo1} when studying the
Kerr-de Sitter metrics with equal angular momenta.  The essential
point is that one can write
\be
\sum_{i=1}^p (d\nu_i^2 + \nu_i^2 d\phi_i^2) = d\Omega_{2p-1}^2=
(d\psi + A)^2 + d\Sigma_{p-1}^2\,,\qquad \sum_{i=1}^p \nu_i^2 d\phi_i
=d\psi + A\,,\label{cpdef}
\ee
where $d\Sigma_{p-1}^2$ is the standard Fubini-Study metric on
$\CP^{p-1}$ (with $R_{ab}= 2p g_{ab}$), and $\ft12 dA$ locally
gives the K\"ahler form $J$. Note that $d\Omega_{2p-1}^2$ is the
standard metric on the unit sphere $S^{2p-1}$, expressed here as the
Hopf fibration over $\CP^{p-1}$.

    With these results, and the analogous ones for the tilded
coordinates $\td\nu_j$ and $\td\phi_j$, we find that (\ref{oddmet})
can be rewritten as
\bea
ds^2 &=& - \fft{(1+g^2r^2)(1-g^2v^2)}{\Xi_a\Xi_b} dt^2
+\fft{\rho^{2n-2} dr^2}{U} + \fft{\omega^{2n-2} dv^2}{V}\nn\\
&&
+\fft{2M}{\rho^{2n-2}} 
  \Big(\fft{(1-g^2 v^2)}{\Xi_a\Xi_b}\, dt - \cA\Big)^2 +
\fft{2L}{\omega^{2n-2}} 
  \Big(\fft{(1+g^2 r^2)}{\Xi_a\Xi_b}\, dt - \wtd\cA\Big)^2\nn\\
&&
 +
\fft{(r^2 + a^2)(a^2-v^2)}{\Xi_a (a^2-b^2)}
\Big( (d\psi + A )^2 + d\Sigma_{p-1}^2\Big)\nn\\
&&
+\fft{(r^2 + b^2)(b^2-v^2)}{\Xi_b (b^2-a^2)}
\Big( (d\varphi + \wtd A )^2 + d\wtd\Sigma_{q-1}^2\Big)\,,
\label{cpform}
\eea
now with
\bea
\cA&=&\fft{a(a^2-v^2)}{\Xi_a(a^2-b^2)} (d\psi + A)
+\fft{b(b^2-v^2)}{\Xi_b(b^2-a^2)} (d\varphi + \wtd A)
\nn\\
\wtd\cA&=&\fft{a(r^2+a^2)}{\Xi_a(a^2-b^2)} (d\psi + A)
+\fft{b(r^2+b^2)}{\Xi_b(b^2-a^2)} (d\varphi + \wtd A)\,.
\label{d5complexp}
\eea
Here $A$ and $\wtd A$ are potentials such that the K\"ahler forms of
the complex projective spaces $\CP^{p-1}$ and $\CP^{q-1}$ are given
locally by $J=\ft12 dA$ and $\wtd J=\ft12\wtd A$ respectively.
Another useful way of writing the metric is given in the appendix.

   It can be seen from the form of (\ref{cpform}) that the metrics
have cohomogeneity 2, with principal orbits on the surfaces where $r$
and $v$ are constant that are the homogeneous spaces $\R\times
S^{2p-1}\times S^{2q-1}$.  The $\R$ factor is associated with the time
direction, whilst the spheres $S^{2p-1}$ and $S^{2q-1}$ arise from the
Hopf fibrations over $\CP^{p-1}$ and $\CP^{q-1}$ respectively.  The
sphere metrics on the principal orbits are squashed, and so the
isometry group of (\ref{cpform}) is $\R\times U(p)\times U(q)$.
   
   We have presented the new solutions for the case of negative
cosmological constant, but clearly these NUT generalisations of
Kerr-AdS will also be valid if we send $g\rightarrow \im\, g$,
yielding NUT generalisations of the Kerr-de Sitter metrics.  It is
also worth noting that even when the cosmological constant is set to
zero, the solutions are still new, representing NUT generalisations
of the asymptotically-flat rotating black holes of Myers and Perry 
\cite{myeper}.

   Written in the form (\ref{oddmet}) or (\ref{cpform}), the metric
appears to be singular in the special case where one sets $a=b$.  This
is, however, an artefact of our introduction of the coordinate $v$, in
place of $\theta$.  We did this in order to bring out the symmetrical
relation between $r$ and $v$, but clearly, as can be seen from
(\ref{thetav}), the coordinate $v$ degenerates in the case $a=b$.
This can be avoided by using $\theta$ as the coordinate instead, and
performing appropriate rescalings.

   Having written our new Kerr-AdS-Taub-NUT metrics in this form, it
is clear that we could obtain more general Einstein metrics by
replacing the Fubini-Study metrics $d\Sigma_{p-1}^2$ and
$d\wtd\Sigma_{q-1}^2$ on $\CP^{p-1}$ and $\CP^{q-1}$ by arbitrary
Einstein-K\"ahler metrics of the same dimensions, and normalised to
have the same cosmological constants as $d\Sigma_{p-1}^2$ and
$d\wtd\Sigma_{q-1}^2$.  In the generalised metrics, $A$ and $\wtd A$
will now be potentials yielding the K\"ahler forms of the two
Einstein-K\"ahler metrics, \ie $J=\ft12 dA$ and $\wtd J= \ft12 d\wtd
A$.

   If we specialise to the case when $b=0$, and define a new
coordinate $\psi'=\psi- a g^2 t$, then the metric (\ref{d5complexp})
reduces to
\bea 
ds^2 &=& \frac{r^2 + v^2}{X} dr^2 + \frac{r^2 + v^2}{Y} dv^2
- \frac{X}{r^2 + v^2}\Big(dt - \frac{a^2 - v^2}{a\, \Xi_a}
(d\psi' + A)\Big)^2\nn\\
&& + \frac{Y}{r^2 + v^2}\Big(dt - \frac{r^2 + a^2}{a\,\Xi_a}
(d\psi' + A)\Big)^2 + \frac{(r^2 + a^2)(a^2 - v^2)}{a^2\Xi_a}
d\Sigma^2_{p-1}\nn\\
&& + \frac{r^2 v^2}{a^2}d\Omega^2_{2q-1}\,,\label{b0met}
\eea
where $d\Omega_{2q-1}^2= (d\varphi+\wtd A)^2 + d\wtd\Sigma_{q-1}^2$ is
the metric of a unit sphere $S^{2q-1}$, and
\bea
X &=& (1 + g^2 r^2) (r^2 + a^2) 
-\frac{2 M}{(r^2 + a^2)^{p-1}\, r^{2(q-1)}}\,,\nn\\
Y &=& (1 - g^2 v^2) (a^2 - v^2) -
\frac{2 \wtd L}{(a^2 -v^2)^{p-1}\, v^{2(q-1)}}\,.
\eea
The constant $\wtd L$ is related to the original NUT parameter by
$\wtd L= (-1)^q\, L$.  A special case of the metrics (\ref{b0met}),
namely when $p=1$, was obtained in \cite{klemm,chgilupo}.

\subsection{The supersymmetric limit}

    Odd-dimensional Kerr-AdS black holes admit supersymmetric limits,
which in Euclidean signature with positive cosmological constant
become Einstein-Sasaki metrics \cite{cvlupapo1,cvlupapo2} (see also
\cite{cvgilupo3,gao} for discussions of how the supersymmetric limit
arises in the Lorentzian regime, when a \bog inequality is saturated).
We find that an analogous limit also exists for our new metrics where
the NUT charge is introduced.  We first set $g={\rm i}$ so that the
metric has a unit positive cosmological constant ($R_{\mu\nu} =
(D-1)\, g_{\mu\nu}$). We then Euclideanise the metric by sending
\be
t \rightarrow {\rm i} t\,,\qquad
\qquad a \rightarrow {\rm i} a\,,\qquad b \rightarrow
{\rm i} b\,,
\ee
define
\bea
&&1 - a^2 = \alpha \,\epsilon\,,\qquad 1 - b^2 = \beta
\,\epsilon\,,\qquad M = - m\, \epsilon^{n+1}\,,\qquad L = \ell
\,\epsilon^{n+1}\,,\nn\\
&&1 - r^2 = \epsilon\, x \,,
\qquad 1 + v^2 = \epsilon\, y \,,
\eea
and then take the limit $\epsilon\rightarrow 0$.
This leads to the metric
\bea
ds^2 &=& \Big(dt + \fft{(\alpha-x)(\alpha- y)}{\alpha(\alpha-\beta)}
(d\psi + A)- \fft{(\beta-x)(\beta-y)}{\beta(\alpha-\beta)}
(d\varphi + \wtd A)\Big)^2 \nn\\
&& + \fft{x - y}{4 X} dx^2 + \frac{x - y}{4Y} dy^2 + 
\fft{(\alpha-x)(\alpha-y)}{\alpha(\alpha-\beta)}d\Sigma^2_{p-1} -
\fft{(\beta - x)(\beta - y)}{\beta(\alpha-\beta )}
d\wtd\Sigma^2_{q-1} \nn\\
&& + \fft{X}{x-y}\Big(\fft{(\alpha-y)}{\alpha(\alpha-\beta)}(d\psi + A)
- \fft{(\beta-y)}{\beta(\alpha -
\beta)} (d\varphi + \wtd A)\Big)^2 \nn\\
&&+ \fft{Y}{x - y} \Big(\fft{(\alpha-x)}{\alpha(\alpha-\beta)}(d\psi + A)
- \fft{(\beta-x)}{\beta(\alpha - \beta)} (d\varphi + \wtd A)\Big)^2
\,,\label{2n1susy}
\eea
where again $J=\ft12 dA$ and $\wtd J=\ft12 d\wtd A$ are the K\"ahler
forms of the $\CP^{p-1}$ and $\CP^{q-1}$ complex projective spaces
with metrics $d\Sigma_{p-1}^2$ and $d\wtd\Sigma_{q-1}^2$ respectively,
and
\bea
X &=& - \frac{2 m}{(\alpha-x)^{p-1}(\beta-x)^{q-1}} - x
(\alpha - x)(\beta-x)\,,\nn\\
Y &=&  \frac{2 \ell}{(\alpha-y)^{p-1}(\beta-y)^{q-1}} 
  + y (\alpha - y)(\beta-y)
\,.\label{XYds2n}
\eea
It is straightforward to verify that the above metric (\ref{2n1susy})
is an Einstein-Sasaki metric in $D=2n+1$ dimensions.  Note that the
metric has the form
\be
ds_{2n+1}^2 = (dt + 2\cA)^2 + ds_{2n}^2\,.
\ee
where $ds_{2n}^2$ is an Einstein-K\"ahler metric and $\cA$ is the
corresponding K\"ahler potential, in the sense that the K\"ahler form
for $ds_{2n}^2$ can be written locally as $J=d\cA$.  As far as we
know, these cohomogeneity-2 Einstein-K\"ahler metrics $ds_{2n}^2$ have
not been obtained explicitly before.  Note that one can go to the
Ricci-flat limit of $ds_{2n}^2$ by performing a rescaling that amounts
to dropping the $x^3$ term and $y^3$ term in (\ref{XYds2n}).

    If we consider the special case where $p=n-1$ and $q=1$, the 
Einstein-Sasaki metrics reduce to ones that were obtained recently in
\cite{lupova}.  This may be seen by defining new parameters by the
expressions
\be
\hat\alpha=-4(\beta-2\alpha)\,,\quad \hat\beta=\alpha(\alpha-\beta)\,,
\quad m= \ft18 (-1)^N\, \mu\,,\quad \ell= \ft18 (-1)^N\, \nu\,,
\ee
where $N\equiv p-1$, and introducing new coordinates defined by
\be
 \hat x= x-\alpha\,,\quad \hat y= y-\alpha\,,\quad t=
      \tau + 2(\alpha-\beta)\chi\,,\quad \varphi= 2\beta\chi\,,\quad
 \psi = 2 \hat\psi + 2\alpha \chi\,.
\ee
Defining also $\hat X= 4X$ and $\hat Y=4Y$, we obtain, upon substitution
into (\ref{2n1susy}), the metric
\bea
ds^2 &=& [d\tau -2 (\hat x+ \hat y) d\chi + 
       \fft{2\hat x\hat y}{\hat\beta} \sigma]^2
     + \fft{\hat x-\hat y}{\hat X}\, d\hat x^2
  + \fft{\hat x-\hat y}{\hat Y}\, d\hat y^2\nn\\
&& +
 \fft{\hat X}{\hat x-\hat y}\,
(d\chi- \fft{\hat y}{\hat \beta}\, \sigma)^2 +
  \fft{\hat Y}{\hat x-\hat y}\,
(d\chi - \fft{\hat x}{\hat \beta}\, \sigma)^2 +
\fft{\hat x\hat y}{\hat \beta}\, d\Sigma_N^2\,,
\eea
where $\sigma= d\hat\psi + \ft12 A$, and 
\be
\hat X = - 4 \hat x^3 - \hat\alpha \hat x^2 - 4 \hat\beta \hat x 
  - \fft{\mu}{\hat x^N}\,,\qquad
\hat Y = 4 \hat y^3 + \hat\alpha \hat y^2 + 4\hat\beta \hat y 
    + \fft{\nu}{\hat y^N}\,.\label{XYn2}
\ee
This is precisely of the form of the Einstein-Sasaki metrics that were
obtained in section (4) of reference \cite{lupova}, in the case where
the Einstein-K\"ahler base metric in that paper is taken to be
$\CP^N$.  A detailed discussion of the global structure of these
metrics was given in \cite{lupova}, and new complete $D=7$
Einstein-Sasaki spaces were obtained.

\section{Kerr-de Sitter with NUT Parameter in $D=2n$}

   The Kerr-de Sitter metrics in even spacetime dimensions take a
slightly different form from those in odd dimensions.  The reason for
this is that now there are an odd number of spatial dimensions, and so
there can be $(n-1)$ independent parameters characterising rotations
in $(n-1)$ orthogonal 2-planes, with one additional spatial direction
that is not associated with a rotation.  Because of this feature, the
$D=2n$ dimensional Kerr-de Sitter black holes in general have
cohomogeneity $n$, which can be reduced to cohomogeneity 2 if one sets
all the $(n-1)$ rotation parameters equal.  By contrast, in odd
dimensions $D=2n+1$ the general metrics have cohomogeneity $n$,
reducing to cohomogeneity 1 if one sets all the rotation parameters
equal.

   It will be recalled that in section \ref{oddsec}, we were able to
generalise the odd-dimensional Kerr-de Sitter to include a NUT
parameter by dividing the angular momentum parameters $a_i$ into two
sets, equal within a set, thereby obtaining a metric of cohomogeneity
2.  Our construction with the NUT parameter is intrinsically adapted
to metrics of cohomogeneity 2, and so this means that in the present
case, when we consider generalising the even-dimensional Kerr-de
Sitter metrics, we shall first need to divide the rotation parameters
$a_i$ into two sets.  In one set, the parameters will be equal and
non-zero, while in the other set, the remaining rotation parameters
will all be chosen to be zero.

   Our starting point is the expression for the Kerr-de Sitter metrics
given in equation (E.3) of reference \cite{gilupapo1}, specialised to
dimension $D=2n$.  We shall take the cosmological constant to be
negative, with the resulting Kerr-AdS metrics satisfying $R_{\mu\nu}=
- (D-1)g^2\, g_{\mu\nu}$.  We then set
\be
 a_1=a_2=\cdots = a_p = a\,,\qquad a_{p+1}= a_{p-2}=\cdots
=a_{n-1} = 0 \,.
\label{aeven}
\ee
We then introduce new ``latitude'' coordinates $\nu_i$, $\td \nu_j$ and 
$\theta$, in place of the $\mu_i$ in \cite{gilupapo1},  
\bea 
\mu_i &=& \nu_i\, \sin\theta\,, \qquad 1\le i\le p\,,\qquad
   \sum_{i=1}^p \nu_i^2=1\,,\nn\\
\mu_{j+p} &=& \td\nu_j\, \cos\theta\,,\qquad 1\le j \le n-p\,,\qquad
  \sum_{j=1}^{n-p} \td\nu_j^2=1\,,
\eea
In this case, because there are only $(n-1)$ azimuthal coordinates
$\phi_i$, we split them into two sets, which we shall denote by
$\phi_i$ and $\td\phi_j$, defined for
\be
\phi_i:\qquad 1\le i\le p\,,\qquad \qquad 
\td\phi_j: \qquad 1\le j\le q\,,
\ee
where this time we have defined $q$ such that
\be
p+q = n-1\,.
\ee
We then introduce a new variable $v$, in place of $\theta$, which this
time is defined by
\be
a^2 \cos^2\theta = v^2\,.
\ee

   We can now write out the Kerr-AdS metric of
\cite{gilupapo1,gilupapo2}, subject to the restriction (\ref{aeven}),
in terms of the new variables defined above, and, as in the
odd-dimensional case we discussed previously, this allows us to
conjecture a generalisation that includes a NUT parameter $L$ as well
as the mass parameter $M$ and angular momentum parameter $a$.  Again,
we shall just present our final result, having included the NUT
parameter.  Thus we obtain the new Kerr-AdS-Taub-NUT metric (which we
have verified explicitly in $D\le 8$)
\bea
ds^2 &=& - \fft{(1+g^2r^2)(1-g^2v^2)}{\Xi_a} dt^2
+\fft{\rho^{2n-3} dr^2}{U} + \fft{\omega^{2n-3} dv^2}{V}\nn\\
&&
+\fft{2M\,r}{\rho^{2n-3}} \Big(\fft{(1-g^2 v^2)}{\Xi_a}\, dt -
\cA\Big)^2 -
\fft{2L\,v}{\omega^{2n-3}} \Big(\fft{(1+g^2 r^2)}{\Xi_a}dt - 
   \wtd\cA \Big)^2\label{evenmets}\\
&&
 +
\fft{(r^2 + a^2)(a^2-v^2)}{a^2\Xi_a}
\sum_{i=1}^p (d\nu_i^2 + \nu_i^2 \, d\phi_i^2)
+\fft{r^2 v^2}{a^2}
\Big(d\td v_{q+1}^2 + \sum_{j=1}^{q} (d\td\nu_j^2 + 
                                  \td\nu_j^2 d\td\phi_j^2)\Big)\,,\nn
\eea
where
\bea
\cA&=&\fft{a^2-v^2}{a\,\Xi_a} \sum_{i=1}^p\nu_i^2 d\phi_i\,,\qquad
\wtd\cA=\fft{r^2+a^2}{a\,\Xi_a} \sum_{i=1}^p\nu_i^2 d\phi_i\,,\nn\\
U&=&(1 + g^2 r^2)(r^2 + a^2)^p\,  r^{2q} - 2M\,r\,,
\nn\\
V&=&(1-g^2v^2)(a^2-v^2)^p \, v^{2q} - 2L\,v\,.
\nn\\
\rho^{2n-3}&=& (r^2 + v^2) (r^2 + a^2)^{p-1} \, r^{2q}
\,,\qquad \Xi_a=1-a^2g^2\,,\nn\\
\omega^{2n-3}&=& (r^2 + v^2) (a^2-v^2)^{p-1}\, 
v^{2q}\,.
\eea

    The $(2n)$-dimensional Kerr-AdS-Taub-NUT metrics that we have
constructed here can be seen to be quite similar in structure to the
$(2n+1)$-dimensional examples that we constructed in section
\ref{oddsec}, in the special case where we set the $b$ parameter to
zero.  In fact we can re-express the metrics (\ref{evenmets}) in terms
of a complex projective space and a sphere metric, in a manner that is
closely analogous to (\ref{b0met}).  This is expressed most simply by
making redefinitions as in (\ref{cpdef}), and then introducing a new
Hopf fibre coordinate $\wtd\psi= \psi -a g^2 t$ as we did in the
odd-dimensional case.  Having done this, we arrive at the metric
\bea
ds^2&=&\fft{r^2 +v^2}{X} dr^2 + \fft{r^2 + v^2}{Y} dv^2 -
\fft{X}{r^2 + v^2} \Big(dt - \fft{a^2 - v^2}{a\,\Xi_a} 
 (d\wtd\psi + A)\Big)^2\label{evenmet2} \\
&& + \fft{Y}{r^2 + v^2} 
  \Big(dt - \fft{a^2 + r^2}{a\, \Xi_a} (d\wtd\psi + A)\Big)^2
+ \fft{(a^2 + r^2) (a^2 - v^2)}{a^2\Xi_a}d\Sigma_{p-1}^2 +
\fft{r^2 v^2}{a^2} d\Omega_{2q}^2\,,\nn
\eea
where $d\Omega_{2q}^2$ is the metric on the unit sphere $S^{2q}$,
\bea
X&=& (1 + g^2 r^2) (r^2 + a^2) - \fft{2M\, r}{(r^2 + a^2)^{p-1}\, 
r^{2q}}\,,\nn\\
Y&=&(1-g^2 v^2) (a^2 - v^2) - \fft{2L\,v}{(a^2 - v^2)^{p-1}\, 
v^{2q}}\,,
\eea
and the K\"ahler form $J$ for the $\CP^{p-1}$ metric $d\Sigma_{p-1}^2$ is
given locally by $J=\ft12 dA$.

       For the cases with $q=0$, there can also be a BPS limit of the
solutions, giving rise to Ricci-flat K\"ahler metrics instead of
Einstein-K\"ahler.  To do this, we first Euclideanise the metric by
setting $t\rightarrow{\rm i}\, t$, $a\rightarrow {\rm i}\, a$ and set
$g={\rm i}$. We then take the following limit
\be
1-a^2=\alpha\, \ep\,,\quad
1-r^2=x\,\epsilon\,,\quad
1+v^2=y\,\epsilon\,,\quad
M=\mu\, (-\epsilon)^{p-1}\,,\quad
L={\rm i}\,\nu\,\epsilon^{p-1}\,,
\ee
with $\epsilon\rightarrow 0$.  The metric becomes $ds^2=\epsilon\,
d\td s^2$, where $d\td s^2$ is a Ricci-flat K\"ahler metric,
given by
\bea
d\td s^2 &=& \fft{x-y}{4X} dx^2 + \fft{x-y}{4Y} dy^2 +
\fft{(x-\alpha)(\alpha-y)}{\alpha}
d\Sigma_{p-1}^2\nn\\
&&+\fft{X}{x-y} (dt + \fft{\alpha-y}{\alpha} (d\psi+A))^2
+\fft{Y}{x-y} (dt - \fft{x-\alpha}{\alpha} (d\psi +A))^2
\,,\nn\\
X&=& x(x-\alpha) + \fft{2\mu}{(x-\alpha)^{p-1}}\,,\qquad
Y=y(\alpha-y) - \fft{2\nu}{(\alpha-y)^{p-1}}\,.
\eea
The K\"ahler 2-form is given locally by $J=dB$, where
\be
B=\ft12 (x+y) dt + \fft{(x-\alpha)(\alpha-y)}{2\alpha}
(d\psi + A)\,.
\ee

\section{Global Analysis}

     The global analysis of Kerr-AdS black holes in general dimensions
was given in \cite{gilupapo1,gilupapo2}.  Here, we study the effect of
introducing the NUT charge $L$.  We shall consider the case where $v$
is a compact coordinate, ranging over the interval $0< v_1\le v\le
v_2$, where $v_1$ and $v_2$ are two adjacent roots of $V(v)=0$, such
that the function $V$ is positive when $v$ lies within the interval.
In the case when $L=0$, we would have $v_1=a$ and $v_2=b$.  The
coordinate $r$ ranges from $r_0$ to infinity, where $r_0$ is the
largest root of $U(r)=0$.  The discussion now divides into the cases
of $D=2n+1$ dimensions and $D=2n$ dimensions.

\subsection{ $D=2n+1$ dimensions}

   The metric (\ref{cpform}) 
is degenerate at $v=v_1$ and $v_2$, where $V(v_i)=0$.  The corresponding
Killing vectors whose norms $\ell^2 = g_{\mu\nu}\, \ell^\mu\, \ell^\nu$ 
vanish at these surfaces have the form
\be
\ell=\gamma_0 \fft{\del}{\del t} + \gamma_1 \fft{\del}{\del\phi} +
\gamma_2 \fft{\del}{\del\psi}\,,
\ee
for constants $\gamma_0$, $\gamma_1$ and $\gamma_2$ to be determined.  The
associated ``surface gravities'' are of Euclidean type, in the sense that
\be
\kappa_E^2 = \fft{g^{\mu\nu}\, (\del_\mu\ell^2)\, (\del_\nu\ell^2)}{
                     4 \ell^2}\Big|_{v=v_i}
\ee
is positive.  Thus these degenerations are typical of an azimuthal
coordinate at a spatial origin.  We can scale the coefficients
$\gamma_i$ so that the Euclidean surface gravity is 1, implying that
the Killing vector generates a closed translation with period $2\pi$.
One might conclude that the time coordinate is periodic, since
$\gamma_0$ is non-vanishing.  This is indeed the case for the
solutions in even dimensions. However, in odd dimensions the
$\del/\del t$ term can be removed by making the coordinate
transformation
\be
t=\td t + \fft{\Xi_b(a^2-v_1^2)(b^2-v_2^2)b\, \psi
-\Xi_a (b^2-v_1^2)(b^2-v_2^2) a\,
\varphi}{ab(a^2-b^2)(1-g^2v_1^2)(1-g^2v_2^2)}
\ee
The two Killing vectors whose norms vanish at $v_1$ and $v_2$ are now
given by
\be
\ell_i=\fft{4L}{V'(v_i)}\Big(\fft{b}{b^2-v_i^2} \fft{\del}{\del\varphi}
+\fft{a}{a^2-v_i^2} \fft{\del}{\del\psi}\Big)\,.
\ee
Both Killing vectors have unit Euclidean surface gravity, implying
that they both generate closed $2\pi$ translations. Since it does not
suffer a periodic identification, $\td t$ is perhaps a more natural
choice than $t$ for the time coordinate.

   The metric also degenerates at $r=r_0$, and the corresponding null
Killing vector has Lorentzian surface gravity $\kappa$, in the sense
that $\kappa^2=-\kappa_E^2$ is positive.  Thus $r=r_0$ is an horizon.
If we write the null Killing vector in terms of coordinate $\td t$,
normalised to
\be
\td\ell_0=\fft{\del}{\del \td t} + \td\gamma_1 \fft{\del}{\del\phi} +
\td\gamma_2 \fft{\del}{\del\psi}\,,
\ee
where $\td\gamma_1$ and $\td\gamma_2$ are determined from the
condition that $\td\ell_0^2=0$ at $r=r_0$, we find that the surface
gravity is given by
\be
\kappa= \fft{(1-g^2 v_1^2)(1-g^2v_2^2)(r_0^2+a^2)(r_0^2 + b^2)
(1 + g^2 r_0^2) U'(r_0)}{2\Xi_a\Xi_b (r_0^2 + v_1^2)(r_0^2 + v_2^2)
[U(r_0)+2M]}\,.
\ee
If instead we consider the null Killing vector in terms of the 
original coordinate $t$, and rescale it to give 
\be
\ell_0=\fft{\del}{\del t} + \gamma_1 \fft{\del}{\del\phi} +
\gamma_2 \fft{\del}{\del\psi}\,,
\ee
then the surface gravity is then given by
\be
\kappa= \fft{(1 + g^2 r_0^2) U'(r_0)}{2[U(r_0)+2M]}\,,
\ee
which is identical to the result for the Kerr-AdS black hole
\cite{gilupapo1,gilupapo2}} without the NUT parameter.  It is not {\it
a priori} obvious what the proper normalisation for the asymptotically
timelike Killing vector should be, since the metrics with the
non-vanishing NUT parameter are not asymptotic to AdS.

\subsection{$D=2n$ dimensions}

         In even dimensions, the introduction of the NUT parameter
implies that the time coordinate is necessarily periodic (as in four
dimensions).  To see this, we note from the metric (\ref{evenmet2})
that, at the degenerate surfaces $v=v_1$ and $v_2$, the Killing
vectors whose norms vanish are given by
\be
\ell_i=\fft{2}{V'(v_i)}\Big( (a^2-v_i^2) \fft{\del}{\del t} +
 a \, \Xi_a\, \fft{\del}{\del\wtd\psi}\Big)\,.
\ee
These Killing vectors are normalised to have unit Euclidean surface
gravities, and hence they generate closed translations with period
$2\pi$. In the case when $L=0$, then $v_1=a$ and $v_2=-a$, so the
$\ell_i$ do not have $\del/\del t$ terms. However, when $L\ne0$ there are
necessarily $\del/\del t$ terms appearing in these Killing vectors that
generate periodic translations, and so $t$ must be identified periodically.

\section{Inversion Symmetry of $D=5$ Kerr-AdS Black Holes}

    In this section, we first demonstrate that the NUT parameter $L$
introduced in our general rotating black holes is trivial in the
special case of $D=5$ dimensions.  However, our demonstration also
brings to light a rather remarkable property of the five-dimensional
Kerr-AdS black hole metric, namely, that it admits a discrete symmetry
transformation which shows that the metric with over-rotation (where
the parameters $a$ and $b$ are such that $a^2 g^2>1$ and/or $b^2
g^2>1$) is equivalent to a Kerr-AdS metric with under-rotation.
 
  We start with the five-dimensional Kerr-AdS metric written in
the (\ref{cpform}) with $p=1$ and $q=1$,
and make the coordinate transformations
\bea
&&\psi\rightarrow a b^2 \chi + a g^2 t + a(1+b^2 g^2) \phi\,,\qquad
\varphi\rightarrow b a^2  \chi + b g^2 t + b(1+a^2 g^2) \phi\,,\nn\\
&&
t\rightarrow t+a^2 b^2 \chi + (a^2+b^2)\phi\,,\label{angdefs}
\eea
and define $r^2 = x$ and $v^2=y$. This leads to the five-dimensional
metric
\bea
ds^2 &=& (x+y)\Big(\fft{dx^2}{4X} + \fft{dy^2}{4Y}\Big) -
\fft{X}{x(x+y)} (dt + y\, d\phi)^2 +
\fft{Y}{y(x+y)} (dt - x\, d\phi)^2\nn\\
&& \fft{a^2b^2}{xy} \Big(dt - xy d\chi -
       (x-y) d\phi\Big)^2\,,\label{5met}
\eea
where
\bea
X &=& (1+g^2 x)(x+a^2)(x+b^2) -2Mx\nn\\
&=& g^2x^3 + (1 + (a^2 + b^2)g^2) x^2 +
(a^2 + b^2 + a^2 b^2 g^2 - 2M) x + a^2b^2\,,\nn\\
Y&=& -(1-g^2 y)(a^2-y)(b^2-y) + 2L y\nn\\
&=& g^2 y^3 - (1 + (a^2 + b^2)g^2) y^2 +
(a^2 + b^2 + a^2 b^2 g^2 + 2L) y - a^2 b^2\,.\label{XYdefs}
\eea

   Although, the solution ostensibly has the four independent
parameters $(M, L, a, b)$, one can in fact scale away either $M$ or
$L$ in this five-dimensional case.  To do this, we set
\be
\td x = \lambda^2 x\,,\quad
\td y = \lambda^2 y\,,\quad
\td t = \fft{t}{\lambda}\,,\quad
\td \chi = \fft{\chi}{\lambda^5}\,,\quad
\td \phi = \fft{\phi}{\lambda^3}\,.
\ee
The metric (\ref{5met}) is invariant under this transformation, if we
simultaneously transform the parameters $a$, $b$, $M$ and $L$. Thus we
define $\wtd X=\lambda^6\, X$ and $\wtd Y=\lambda^6\,Y$, where $\wtd
X$ and $\wtd Y$ are defined as in (\ref{XYdefs}) except with tilded
parameters $\td a$, $\td b$, $\wtd M$ and $\wtd L$.  It follows that
we shall have
\bea
&& \lambda^2 + \lambda^2 (a^2+b^2) g^2 = 1 +
(\td a^2 + \td b^2) g^2\,,\qquad
\lambda^6 \, a^2 \, b^2 = \td a^2\, \td b^2\,,\nn\\
&&\lambda^4(a^2+b^2+ a^2 b^2 g^2 +2L) = 
\td a^2 +\td b^2 + \td a^2 \td b^2 g^2 +
2\wtd L\,,\nn\\
&&\lambda^4(a^2+b^2+ a^2 b^2 g^2 -2M) = 
\td a^2 +\td b^2 + \td a^2 \td b^2 g^2 -
2\wtd M\,.\label{4eqs}
\eea
We can then choose, for example, to set $\wtd L=0$, and solve the four
equations (\ref{4eqs}) for $\td a$, $\td b$, $\wtd M$ and $\lambda$.
Thus a solution with $L\ne0$ is transformed into a tilded solution
with $\wtd L=0$, and since this latter solution is just of the
original five-dimensional Kerr-AdS form, it follows that the metric
(\ref{5met}), even with $L\ne 0$, is also just the five-dimensional
Kerr-AdS metric, but with changed values for the rotation and mass 
parameters.  It is nevertheless interesting that the Kerr-de
Sitter black hole in $D=5$ can be put in such a symmetric form.

   It should be stressed that the scaling symmetry that we used above
in order to show that the parameter $L$ in the five-dimensional
metrics is ``trivial'' is very specific to five dimensions.  In
particular, it can be seen from (\ref{cpform}) that in higher
dimensions, when at least one of $p$ or $q$ exceeds 1, the associated
metrics on the complex projective spaces will break the scaling
symmetry.  Thus, as in the case of the simpler NUT generalisations
discussed \cite{klemm,chgilupo}, five-dimensions is the exception in
not admitting a non-trivial generalisation.

   The transformation described above becomes particularly simple if
we consider the case of an asymptotically flat five-dimensional
rotating black hole, \ie when $g=0$.  In this case, we have from
(\ref{4eqs}) that $\lambda=1$ and
\be
\td a^2 + \td b^2 + 2\wtd L  = a^2 + b^2 + 2L\,,\quad
\td a^2 + \td b^2 -2 \wtd M = a^2 +b^2 -2M\,,\quad
 \td a^2 \td b^2 = a^2 b^2\,.
\ee
Thus $\wtd L + \wtd M=L+M$, and so we can arrange to have $\wtd L=0$
by taking $\wtd M=L+M$, implying that $\td a^2 + \td b^2 = a^2 + b^2 + 2L$,
together with $\td a^2 \td b^2 =a^2 b^2$.  It is worth noting, however,
that even though one can always map into a solution where $\wtd L=0$, 
it may, depending upon the original values for $a$, $b$ and $L$,  
correspond to having complex values for $\td a$ and $\td b$.  Although
the metric (\ref{5met}) would still be real, the metric written back in
terms of the original $\psi$, $\phi$ and $t$ coordinates would then be
complex. Thus although the parameter $L$ is really trivial in five 
dimensions, its inclusion can nevertheless allow one to parameterise
the solutions in a wider class without the need for complex coordinate
transformations.  Similar remarks apply also to the case when $g\ne0$.

   There is another interesting consequence of the five-dimensional
scaling symmetry discussed above, namely, that even with the parameter
$L$ omitted entirely, the five-dimensional rotating AdS black hole
metrics have a symmetry that allows one to map an ``over-rotating''
black hole (\ie where $a^2 g^2>1$ or $b^2 g^2 >1$) into an
under-rotating black hole.  This can be understood by again
considering the transformations in (\ref{4eqs}), where we now choose
not only $\wtd L=0$ but also $L=0$.  The system of equations then
admits a sextet of solutions for $(\td a, \td b, \wtd M, \lambda)$
(where we assume, without loss of generality, that the signs of the
rotation parameters are unchanged):
\bea
\td a &=& a\,,\quad \td b=b\,,\quad \wtd M= M\,,\quad \lambda=1\,,\nn\\
\td a &=& b\,,\quad \td b=a\,,\quad \wtd M= M\,,\quad \lambda=1\,,\nn\\
\td a &=& \fft1{a g^2}\,,\quad \td b= \fft{b}{ag}\,,\quad 
  \wtd M= \fft{M}{a^4 g^4}\,,\quad \lambda= \fft1{ag}\,,\nn\\
\td a &=& \fft1{b g^2}\,,\quad \td b= \fft{a}{bg}\,,\quad 
  \wtd M= \fft{M}{b^4 g^4}\,,\quad \lambda= \fft1{bg}\,,\nn\\
\td a &=& \fft{a}{bg}\,,\quad \td b= \fft1{b g^2}\,,\quad
   \wtd M= \fft{M}{b^4 g^4}\,,\quad \lambda= \fft1{bg}\,,\nn\\
\td a &=& \fft{b}{ag}\,,\quad \td b= \fft1{a g^2}\,,\quad
   \wtd M= \fft{M}{a^4 g^4}\,,\quad \lambda= \fft1{ag}\,.\label{kads5tran}
\eea
The first of these is the identity, the second is merely an exchange
of the r\^oles of $a$ and $b$, whilst the remaining four, modulo
exchanges of the $a$'s and the $b$'s, are equivalent and non-trivial.
Taking the third as an example, we see that if the metric is
over-rotating by virtue of having $a^2g^2>1$, then it can be
re-expressed, by a change of variables, as a metric which is
under-rotating.  In fact any five-dimensional Kerr-AdS black hole with
over-rotation is equivalent, after a change of coordinates, to one
with under-rotation.  Of course, after transforming back into the
original coordinates in which the over-rotating black hole ostensibly
exhibited singular behaviour, one would find that the coordinate
ranges that actually reveal that it is well-behaved are not the
``naive'' ones that led to the original conclusion of singular
behaviour.

   It is instructive to rewrite the transformations (\ref{kads5tran})
in terms of the original coordinates of the five-dimensional 
Kerr-AdS metric as given by Hawking, Hunter and Taylor-Robinson in
\cite{hawhuntay}.  The metric is given by
\bea
ds_5^2 &=& -\fft{\Delta}{\rho^2}\, \Big[ dt -
   \fft{a\, \sin^2\theta}{\Xi_a}\, d\phi - \fft{b\, \cos^2\theta}{\Xi_b}\,
d\psi\Big]^2 + \fft{\Delta_\theta\, \sin^2\theta}{\rho^2}\,
\Big[ a\, dt -\fft{r^2+a^2}{\Xi_a}\, d\phi\Big]^2\nn\\
&&  \fft{\Delta_\theta\, \cos^2\theta}{\rho^2}\,
\Big[ b\, dt -\fft{r^2+b^2}{\Xi_b}\, d\psi\Big]^2 +
\fft{\rho^2\, dr^2}{\Delta} + \fft{\rho^2\, d\theta^2}{\Delta_\theta}
\nn\\
&& + \fft{(1+ g^2 r^2)}{r^2\, \rho^2}\,
\Big[ a\, b\, dt - \fft{b\, (r^2+a^2)\, \sin^2\theta}{\Xi_a}\, d\phi
- \fft{a\, (r^2+b^2)\, \cos^2\theta}{\Xi_b}\, d\psi\Big]^2\,,
\label{hawkmet}
\eea
where
\bea
\Delta &\equiv & \fft1{r^2}\, (r^2+a^2)(r^2+b^2)(1 + g^2r^2) -2M\,,\nn\\
\Delta_\theta &\equiv& 1 -a^2\, g^2\, \cos^2\theta -
           b^2\, g^2\, \sin^2\theta\,,\nn\\
\rho^2 &\equiv& r^2 + a^2\, \cos^2\theta + b^2\, \sin^2\theta\,,\nn\\
\Xi_a &\equiv& 1 - a^2\, g^2\,,\qquad \Xi_b \equiv 1- b^2\, g^2\,.
\eea
It satisfies $R_{\mu\nu}=-4 g^2\, g_{\mu\nu}$.  Taking the
transformation in the third line of (\ref{kads5tran}) as an example,
we find that after re-expressing our results back in terms of the
quantities in (\ref{hawkmet}), the symmetry transformation amounts to
\bea
&& a\rightarrow \fft1{a g^2}\,,\quad b\rightarrow \fft{b}{a g}\,,\quad
M\rightarrow \fft{M}{a^4 g^4}\,,\nn\\
&&
\phi\rightarrow -\fft{1}{ag}\, \phi\,,\quad
\psi\rightarrow \psi - \fft{b}{a}\, \phi\,,\quad
t \rightarrow a g t + \fft1{g}\, \phi\,,\nn\\
&& r \rightarrow \fft1{ag}\, r\,,\quad 
  \cos\theta \rightarrow \Big(1 - \fft{\Xi_a}{\Xi_b}\Big)^{1/2}\,
\cos\theta\,.
\label{hawtaytran}
\eea
It is straightforward to see that this transformation leaves the
metric in (\ref{hawkmet}) invariant, and that it therefore allows one
to map an over-rotating Kerr-AdS metric into an under-rotating one.
In other words, if we perform the transformation of parameters given
in the first line in (\ref{hawtaytran}), then the metric is restored
to its original form by making the general coordinate transformations
given also in (\ref{hawtaytran}). 

    Another way of expressing this result is that for any given values of
$a$ and $b$, and provided one allows the coordinates to take complex
values in general, then there exist real sections of the complex metric
describing Kerr-AdS black holes with under-rotation, and also
real sections of the same metric that describe Kerr-AdS black holes with  
over-rotation.

   It is instructive also to re-express the coordinate transformations
in (\ref{hawtaytran}) in terms of the coordinates $y$ and $\hat\theta$
rather than $r$ and $\theta$, where $y$ and $\hat\theta$ are the
coordinates with respect to which the conformal boundary of the
Kerr-AdS metric is precisely the standard $\R\times S^3$ Einstein
universe, with a round $S^3$ factor.  They are defined by
\cite{hawhuntay}
\be
\Xi_a\, y^2\, \sin^2\hat\theta = (r^2+a^2)\, \sin^2\theta\,,\qquad
\Xi_b\, y^2\, \cos^2\hat\theta = (r^2+b^2)\, \cos^2\theta\,.
\ee
Applying the transformations in (\ref{hawtaytran}), we find that these
imply the coordinate transformations
\be
y^2\rightarrow - \fft1{g^2} - y^2 \sin^2\hat\theta\,,\qquad
\tan^2\hat\theta \rightarrow -\Big(1 +\fft{1}{g^2 y^2}\Big)\, 
      \sec^2\hat\theta\,.\label{ythetatran}
\ee
This result emphasises that the original $y=$constant boundary, which
is the most natural choice from the AdS/CFT point of view
\cite{hawhuntay,gipepo3}, is quite different from the $y=$constant
boundary of the transformed metric.

   A number of remarks are in order.  First, we note that the symmetry
we are discussing, which can be expressed in terms of dimensionless
quantities as $ag\rightarrow 1/(ag)$, exists only in the case of the
rotating black hole with a cosmological constant.  In the case of
asymptotically-flat black holes, for which $g=0$, there is no
inversion symmetry.  The inversion symmetry for the five-dimensional
Kerr-AdS black hole is reminiscent of a T-duality symmetry, in the
sense that it implies there is a maximum allowed value for the
rotation, namely $a^2g^2=1$.  In fact, this value is associated with
the supersymmetric limit.  If one considers the case where a rotation
parameter is becoming very large, \ie $a^2g^2 >>1$, then it can be
seen from (\ref{hawtaytran}) that in the limiting case when $a^2g^2$
approaches infinity, the metric will actually approach the pure AdS
metric.

   It is interesting also to consider the effect on the canonical AdS
metric of the transformations (\ref{ythetatran}) taken in isolation.
In other words, we start with the AdS metric
\be
ds^2 = -(1+g^2 y^2) dt^2 + \fft{dy^2}{1+g^2 y^2} + y^2(
   d\hat\theta^2 + \sin^2\hat\theta^2 \,d\phi^2 + \cos^2\hat\theta\,
   d\psi^2)\,,\label{pureads}
\ee
and impose just the coordinate transformations given in
(\ref{ythetatran}) (which are independent of the rotation parameters
$a$ and $b$).  Upon doing so, we find that the AdS metric
(\ref{pureads}) transforms according to
\be
ds^2 \rightarrow  -\fft1{g^2}\, (1+g^2 y^2) d\phi^2 + 
\fft{dy^2}{1+g^2 y^2} + y^2(
   d\hat\theta^2 + \sin^2\hat\theta^2 \,g^2 dt^2 + \cos^2\hat\theta\,
   d\psi^2)\,.
\ee
This is identical in form to (\ref{pureads}), with the r\^oles of
$\phi$ and $gt$ exchanged.  It can easily be seen that in terms of the
standard embeddding of AdS$_5$ in $\R^{4,2}$, the transformation
(\ref{ythetatran}) corresponds to exchanging the the r\^oles of the
two timelike embedding coordinates with a pair of spacelike embedding
coordinates.

\section{Conclusions}

   In this paper, we have constructed generalisations of certain
Kerr-de Sitter and Kerr-AdS black holes in all dimensions $D\ge 6$, in
which an additional NUT-type parameter is introduced.  Specifically,
the cases where we have obtained the more general solutions are where
the rotation parameters are specialised so that the metrics have
cohomogeneity 2.  The nature of the generalisation is then analogous
to the way in which a NUT parameter can be introduced in the
four-dimensional Kerr-de Sitter metrics.

   The same procedure can be followed also in five dimensions, but in
this case we find that the additional NUT parameter is trivial, in the
sense that it can be absorbed by a rescaling of parameters and
coordinates.  However, we also found that there exists a remarkable
symmetry of the five-dimensional Kerr-AdS metrics, in which one can
map a solution where one or both of the rotation parameters are large
(the case of over-rotation, where $a^2g^2>1$ and/or $b^2 g^2>1$) into
a solution where the rotation parameters are small (\ie
under-rotation).  This means that there is effectively a maximum
rotation possible, corresponding to the supersymmetric case where
$a^2g^2=1$ or $b^2 g^2=1$.

   We also studied the supersymmetric limits of the new Kerr-de
Sitter-Taub-NUT metrics, showing that after Euclideanisation we can
obtain new cohomogeneity-2 Einstein-Sasaki metrics in all odd
dimensions $D\ge7$, and new cohomogeneity-2 Ricci-flat K\"ahler
metrics in all even dimensions $D\ge6$.

\section*{Acknowledgements}

   We than Zhiwei Chong and Justin V\'azquez-Poritz for helpful
discussions.

\newpage

\centerline{{\Large{\bf Appendix}}}

\section*{Another Form for the Odd-Dimensional Metrics}

   If we perform the same angular redefinitions (\ref{angdefs}) in the
general odd-dimensional metrics (\ref{cpform}), they may be
re-expressed as
\bea
ds^2 &=&\fft{r^2+v^2}{X}\, dr^2 + \fft{r^2+v^2}{Y}\,dv^2 + 
  \fft{(r^2+a^2)(a^2-v^2)}{\X_a (a^2-b^2)}\, d\Sigma_{p-1}^2 +
   \fft{(r^2+b^2)(b^2-v^2)}{\Xi_b (b^2-a^2)}\, d\wtd\Sigma_{q-1}^2\nn\\
&& \!\!+ \fft{a^2b^2}{r^2v^2}\, \Big[ dt - (r^2-v^2)d\phi - r^2 v^2 d\chi -
       \fft{(r^2+a^2)(a^2-v^2)}{a \Xi_a (a^2-b^2)}\, A -
       \fft{(r^2+b^2)(b^2-v^2)}{b\Xi_b(b^2-a^2)}\, B\Big]^2\nn\\
&& - \fft{X}{r^2+v^2}\, \Big[ dt + v^2 d\phi - 
        \fft{a(a^2-v^2)}{\Xi_a(a^2-b^2)}\, A - 
            \fft{b(b^2-v^2)}{\Xi_b (b^2-a^2)}\, B\Big]^2\nn\\
&&+ \fft{Y}{r^2+v^2}\, \Big[ dt - r^2 d\phi - 
        \fft{a(r^2+a^2)}{\Xi_a(a^2-b^2)}\, A -
         \fft{b(r^2+b^2)}{\Xi_b (b^2-a^2)}\, B\Big]^2\,,
\eea
where we have defined $X$ and $Y$ as
\bea
X&\equiv& \fft{U}{(r^2+a^2)^{p-1}\, (r^2+b^2)^{q-1}}\nn\\
&=& 
  \fft{(1+g^2 r^2)(r^2+a^2)(r^2+b^2)}{r^2} - \fft{2M}{(r^2+a^2)^{p-1}\, 
                (r^2+b^2)^{q-1}}\,,\nn\\
Y&\equiv & \fft{V}{(a^2-v^2)^{p-1}\, (b^2-v^2)^{q-1}}\nn\\
& =&
   \fft{-(1-g^2 v^2)(a^2-v^2)(b^2-v^2)}{v^2} + 
  \fft{2L}{(a^2-v^2)^{p-1}\, (b^2-v^2)^{q-1}}\,.
\eea
This form can sometimes be useful, since it is expressed in a 
manifest ``vielbein basis.''

\end{document}